\documentstyle[11pt,amssymb]{article}

\input psfig.sty

\let\a=\alpha    
    
  \let\n=\nu

\let\C=\Chi

\def\nn{\nonumber} \def\bd{\begin{document}} \def\ed{\end{document}}
\def\ds{\documentstyle} \let\fr=\frac \let\bl=\bigl \let\br=\bigr
\let\Br=\Bigr \let\Bl=\Bigl 
\let\bm=\bibitem
\let\na=\nabla
\let\pa=\partial \let\ov=\overline 
\newcommand{\be}{\begin{equation}} 
\newcommand{\ee}{\end{equation}} 
\def\ba{\begin{array}}
\def\ea{\end{array}}
\def\ft#1#2{{\textstyle{{\scriptstyle #1}\over {\scriptstyle #2}}}}
\def\fft#1#2{{#1 \over #2}}
\def\del{\partial}
\def\vp{\varphi}
\def\st#1{{\scriptstyle #1}}
\def\sst#1{{\scriptscriptstyle #1}}

\def\oneone{\rlap 1\mkern4mu{\rm l}}
\def\td{\tilde}
\def\wtd{\widetilde}
\def\ie{\rm i.e.\ }
\def\dalemb#1#2{{\vbox{\hrule height .#2pt
        \hbox{\vrule width.#2pt height#1pt \kern#1pt
                \vrule width.#2pt}
        \hrule height.#2pt}}}
\def\square{\mathord{\dalemb{6.8}{7}\hbox{\hskip1pt}}}

\def\cramp{\medmuskip = 2mu plus 1mu minus 2mu}
\def\cramper{\medmuskip = 2mu plus 1mu minus 2mu}
\def\crampest{\medmuskip = 1mu plus 1mu minus 1mu}
\def\uncramp{\medmuskip = 4mu plus 2mu minus 4mu}

\newcommand{\ho}[1]{$\, ^{#1}$}
\newcommand{\hoch}[1]{$\, ^{#1}$}
\newcommand{\bea}{\begin{eqnarray}} 
\newcommand{\eea}{\end{eqnarray}} 
\newcommand{\ra}{\rightarrow}
\newcommand{\lra}{\longrightarrow}
\newcommand{\Lra}{\Leftrightarrow}
\newcommand{\ap}{\alpha^\prime}
\newcommand{\bp}{\tilde \beta^\prime}
\newcommand{\tr}{{\rm tr} }
\newcommand{\Tr}{{\rm Tr} } 
\def\0{{\sst{(0)}}}
\def\1{{\sst{(1)}}}
\def\2{{\sst{(2)}}}
\def\3{{\sst{(3)}}}
\def\4{{\sst{(4)}}}
\def\5{{\sst{(5)}}}
\def\6{{\sst{(6)}}}
\def\7{{\sst{(7)}}}
\def\8{{\sst{(8)}}}
\def\n{{\sst{(n)}}}
\def\cA{{{\cal A}}}
\def\cF{{{\cal F}}}
\def\tV{\widetilde V}
\def\tW{\widetilde W}
\def\tH{\widetilde H}
\def\tE{\widetilde E}
\def\tF{\widetilde F}
\def\tA{\widetilde A}
\def\im{{{\rm i}}}
\def\jm{{{\rm j}}}
\def\km{{{\rm k}}}

\def\tY{{{\wtd Y}}}
\def\ep{{\epsilon}}
\def\vep{{\varepsilon}}
\def\R{\rlap{\rm I}\mkern3mu{\rm R}}
\def\bD{{{\bar D}}}
\def\R{{{\Bbb R}}}
\def\C{{{\Bbb C}}}
\def\H{{{\Bbb H}}}
\def\CP{{{\Bbb C}{\Bbb P}}}
\def\RP{{{\Bbb R}{\Bbb P}}}
\def\Z{{{\Bbb Z}}}
\def\bA{{{\Bbb A}}}
\def\bB{{{\Bbb B}}}

\newcommand{\NP}{Nucl. Phys. }
\newcommand{\tamphys}{\it Center for Theoretical Physics\\
Texas A\&M University, College Station, TX 77843, USA}
\newcommand{\umich}{\it Michigan Center for Theoretical Physics\\
University of Michigan, Ann Arbor, Michigan 48109, USA}
\newcommand{\upenn}{\it Department of Physics and Astronomy\\
University of Pennsylvania, Philadelphia,  PA 19104, USA}
\newcommand{\SISSA}{\it  SISSA-ISAS and INFN, Sezione di Trieste\\
Via Beirut 2-4, I-34013, Trieste, Italy}

\newcommand{\ihp}{\it Institut Henri Poincar\'e\\
  11 rue Pierre et Marie Curie, F 75231 Paris Cedex 05}

\newcommand{\damtp}{\it DAMTP, Centre for Mathematical Sciences,
 Cambridge University, Wilberforce Road, Cambridge CB3 OWA, UK}

\newcommand{\auth}{M. Cveti\v{c}\hoch{\dagger 1}, 
H. L\"u\hoch{\star 2} and C.N. Pope\hoch{\ddagger 3}}

\begin{document}
\begin{flushright}
\hfill{{CTP TAMU-14/01}\ \ \ {UPR-936-T}\ \ \
{MCTP-01-21}\\ 
{May 2001}\\
{hep-th/0105096}}
\end{flushright}


\begin{center}
{ \large {\bf Massless 3-branes in M-theory}}

\vspace{8pt}
\auth

\vspace{5pt}
{\hoch{\dagger}\upenn}

\vspace{4pt}
{\hoch{\star}\umich}

\vspace{4pt}
{\hoch{\ddagger}\tamphys}


\vspace{40pt}

\underline{ABSTRACT}
\end{center}

    We construct supersymmetric M3-brane solutions in $D=11$ supergravity.
They can be viewed as deformations of backgrounds taking the form of a
direct product of four-dimensional Minkowski spacetime and a
non-compact Ricci-flat manifold of $G_2$ holonomy.  Although the
4-form field strength is turned on it carries no charge, and the
3-branes are correspondingly massless. We also obtain 3-branes of a
different type, arising as M5-branes wrapped over $S^2$.

{\vfill\leftline{}\vfill
\vskip 5pt
\footnoterule
{\footnotesize \hoch{1} Research supported in part by DOE grant
DE-FG02-95ER40893 and NATO grant 976951. \vskip -12pt} \vskip 14pt
{\footnotesize \hoch{2} Research supported in full by DOE grant
DE-FG02-95ER40899 \vskip -12pt} \vskip 14pt
{\footnotesize  \hoch{3} Research supported in part by DOE
grant DE-FG03-95ER40917.\vskip  -12pt}}

\pagebreak
\setcounter{page}{1}

\vfill\eject
\section{Introduction}

      The standard D3-brane provides a natural supergravity dual of
four dimensional ${\cal N}=4$ superconformal Yang-Mills theory, {\it
via} the AdS/CFT correspondence \cite{malda,gkp,wit}.  Branes with
less supersymmetry can in general be constructed by replacing the
spheres that form the level surfaces in the flat transverse space by
some other Einstein space that admits a lesser number of Killing
spinors \cite{dlps}.  It was proposed that D3-branes on the
six-dimensional conifold, in which the level surfaces are the
$T^{1,1}$ space, is dual to an ${\cal N}=1$ superconformal theory in
$D=4$ with gauge group $SU(N)\times SU(N)$ \cite{klebwit}. The
conformal symmetry can then itself be broken, by introducing
fractional branes corresponding to the wrapping of D5-branes on
2-cycles.  The corresponding supergravity solutions were obtained in
\cite{klebtsey,klebstras}.

     It has been proposed that M-theory compactified on a certain
singular seven-dimensional space with $G_2$ holonomy might be related
to a ${\cal N}=1$, $D=4$ gauge theory \cite{acharya,amv,wit-talk},
which has no conformal symmetry to begin with.  (See also the recent
papers \cite{gomis,edelnun,kacmcg,gutpap,kkp}.)  This leads to the
question of whether there might exist a 3-brane configuration in
M-theory, whose transverse space is a deformation of a Ricci-flat
space of $G_2$ holonomy, in which the 4-form field is turned on.  In
this paper we shall indeed obtain 3-brane solutions of this deformed
type.

   So far, three explicit metrics for seven-dimensional manifolds of
$G_2$ holonomy are known \cite{brysal,gibpagpop}.  They all have
cohomogeneity one.  The first two have principal orbits that are
$\CP^3$ or $SU(3)/(U(1)\times U(1))$, written as an $S^2$ bundle over
$S^4$ or $\CP^2$ respectively.  The associated 7-manifolds have the
topology of $\R^3$ bundles over $S^4$ or $\CP^2$.  The third manifold
has principal orbits that are topologically $S^3\times S^3$, written
as an $S^3$ bundle over $S^3$, and the 7-manifold is topologically
$\R^4\times S^3$.  In order to construct a non-trivial 3-brane
configuration on such a background in eleven-dimensional supergravity,
it is necessary that the background $G_2$ manifold itself should admit
a well-behaved harmonic 4-form (or dual 3-form).  It was shown in
\cite{clptrans,d2ns2} that such harmonic forms exist in all three of
these explicit examples.  In this paper, we construct M3-brane
configurations describing deformations away from backgrounds having a
$G_2$ manifold as the transverse space, taking the 4-form field
strength of M-theory to be proportional to the appropriately deformed
harmonic 4-form.  We first obtain the second-order equations for the
fields in our ansatz, which follow from those of eleven-dimensional
supergravity, and then we show that in a Lagrangian formulation of
these equations the potential can be derived from a superpotential.
This leads to first-order equations which we are able to solve
explicitly.
   
   The exact solutions that we obtain by this method describe 
configurations with a four-dimensional Poincar\'e invariance in the
world-volume, and a seven-dimensional transverse space that is a
deformation of the original Ricci-flat metric of $G_2$ holonomy.  We
may thus view them as being 3-brane solutions of M-theory.  At large
distance they approach the product of 4-dimensional Minkowski
spacetime and the Ricci-flat metric of $G_2$ holonomy.  The rate at
which the metrics approach this asymptotic form is rapid enough that
the ADM mass vanishes, and so they may be thought of as massless
M3-branes.  In common with other examples of massless branes, they
have naked singularities at short distance.  We show that the M3-brane
solutions are supersymmetric.

      It is of interest also to look for 3-brane configurations in
M-theory within a more general framework.  Another natural candidate
for a 3-brane is to look for an M5-brane wrapped on a supersymmetric
2-cycle.  Wrapped supersymmetric M5-branes have been discussed in
previous papers \cite{maldnunez1,gaukimwald}, and typically these have
been of the form AdS$_d\times H_{7-d}$, where $H_n$ denotes the
$n$-dimensional hyperbolic space.  In section 5, we shall consider
M5-branes wrapping around a 2-sphere.  The solutions can be obtained
by starting with $SU(2)$-gauged AdS supergravity in $D=7$, and looking
for 3-branes supported by the Yang-Mills fields.  We obtain the
equations of motion for the general non-abelian case, and show that
when only a $U(1)$ subgroup is turned on, we can construct
first-order equations derivable from a superpotential.  The general
solution can be reduced to Abel's equation, and the structure of the
resulting configurations can be analysed.  The solutions can be lifted
to $D=11$, where they describe 3-branes as M5-branes wrapped on
$S^2$.

\section{M3-branes in backgrounds of $\R^3$ bundles over $S^4$ or $\CP^2$}

    In this section we shall construct M3-brane solutions that can be
viewed as living in backgrounds where the 7-dimensional transverse
space is a manifold of $G_2$ holonomy with the topology of the $\R^3$
bundle over $S^4$ or $\CP^2$.

\subsection{The ansatz}

    Let us consider the $D=11$ ansatz
\be
ds_{11}^2 = H^2\, dx^\mu\, dx_\mu +
   d\rho^2 + a^2\, D\mu^i\, D\mu^i + b^2\, d\Omega_4^2\,,\label{ans1}
\ee
where $\mu^i\,\mu^i=1$ and $D\mu^i=d\mu^i + \epsilon_{ijk}\, A_\1^j\,
\mu^k$, and $A^i_\1$ is the $SU(2)$ Yang-Mills instanton on $S^4$,
whose unit metric is $d\Omega_4^2$.  The functions $H$, $a$ and $b$ will be
taken to depend only on the radial coordinate $\rho$ in the transverse
space.  This describes the case of the $\R^3$ bundle over $S^4$.  The
second possibility is obtained by replacing the $S^4$ by $\CP^2$.
This does not affect the form of the equations for $H$, $a$ and 
$b$.\footnote{In everything that follows, results obtained for the
case of the $S^2$ bundle over $S^4$ apply equally, {\it mutatis
mutandis}, to the case of the $S^2$ bundle over $\CP^2$.}
The constrained $\mu^i$ coordinates can be expressed in terms of two
angular coordinates on $S^2$ in a standard way,
\be
\mu_1 = \sin\theta\, \sin\phi\,,\quad 
\mu_2 = \sin\theta\, \cos\phi\,,\quad
\mu_3 = \cos\theta\,.\label{muang}
\ee
The vielbein components in the $S^2$ fibre directions are then given
by
\bea
e^1 &=& a\, (d\theta - A_\1^1\, \cos\phi + A_\1^2\, \sin\phi)\,,\nn\\
e^2 &=& a\, \sin\theta\, (d\phi + A_\1^1\, \cot\theta\, \sin\phi +
A_\1^2\, \cot\theta\, \cos\phi - A_\1^3)\,.\label{s2viel}
\eea

    There is clearly a vacuum solution of the form (\ref{ans1}) that
is simply the direct product of four-dimensional Minkowski spacetime
and the associated Ricci-flat seven-dimensional manifold with $G_2$
holonomy.  In the vacuum we shall have $H=1$, with $a$ and $b$ being
given by \cite{brysal,gibpagpop}.   We should now like to
turn on the 4-form field strength of eleven-dimensional
supergravity.  The 4-form ansatz that respects the symmetry of the
metric is given by \cite{gibpagpop,d2ns2}
\bea
F_\4 &=& f_1\, \Omega_\4 + f_2\, X_\2\wedge Y_\2 + f_3\, d\rho\wedge
Y_\3\,,\nn\\
{*F}_\4 &=& H^4\, a^2\, b^{-4}\, f_1\, \ep_\4\wedge d\rho\wedge X_\2
   + H^4\, a^{-2}\, f_2\, \ep_\4\wedge d\rho\wedge Y_\2 +
     H^4\, f_3\, \ep_\4\wedge X_\3\,,\label{4form1}
\eea
where the $f_i$ are functions depending only on $\rho$, and 
\bea
&&X_\2\equiv\ft12 \epsilon_{ijk}\,\mu^i\, D\mu^j\wedge D\mu^k\,,\qquad
Y_\2 \equiv \mu^i\, F_\2^i\,,\qquad X_\3\equiv D\mu^i\wedge F_\2^i\,,\nn\\
&&Y_\3\equiv\epsilon_{ijk}\,\mu^i\,D\mu^j\wedge F_\2^k\,,\qquad
\epsilon_\4 \equiv  dt\wedge dx_1\wedge dx_2\wedge dx_3\,.
\eea
(Note that $F_\4$ could in principle have had a term of the form $d\rho\wedge
X_\3$ as well, but this is ruled out by the field equation 
$d{*F_\4}=0$.)  The Bianchi identity $dF_\4=0$
implies that $f_1'=4f_3$ and $f_2'= 2f_3$, so we can
take $f_1=2f$, $f_2=f$ and $f_3=\ft12 f'$, giving
\be
F_\4 = f\, (2\Omega_\4 + X_\2\wedge Y_\2) + \ft12 f'\, d\rho\wedge Y_\3\,.
\label{f4ans}
\ee
In fact $F_\4=dA_\3$ with $A_\3=\ft12 f\, Y_\3$.
The field equation $d{*F_\4}=0$ implies 
\be
(2a^4 + b^4)\, H^4\, f - \ft12 a^2\, b^4\, (H^4\, f')'=0\,.
\ee
(The $F_\4\wedge F_\4$ term vanishes here.)

    In order to impose the $D=11$ Einstein equation it is convenient
to perform a Kaluza-Klein reduction on the 4-dimensional world volume
of the 3-brane, so that the problem can be reformulated from a
seven-dimensional point of view.  This allows us to make use of
curvature calculations for 7-metrics of this type that were performed in
\cite{gibpagpop}.  The relevant seven-dimensional Lagrangian is given by
\be 
e^{-1}{\cal L}_7 = R-\ft12 (\del\phi)^2 -\ft1{48}
e^{\sqrt{\ft85}\phi}\, F_\4^2\,,\label{d7lag1}
\ee
with the ansatz (\ref{ans1}) now taking the form
\bea
ds_7^2 &=& dt^2 + \td a^2\, D\mu^i\, D\mu^i + \td b^2\, d\Omega_4^2\,,
\nn\\
F_\4 &=& f\, (2\Omega_\4 + X_\2\wedge Y_\2) + \ft12 f'\, d\rho\wedge
Y_\3\,.
\eea
The metric in $D=7$ is related to the one in $D=11$ by
\be
d\hat s_{11}^2 = e^{-\fft23\sqrt{\fft25}\,\phi} \, ds_7^2 +
 e^{\fft13\sqrt{\fft52}\,\phi}\, dx^\mu\, dx_\mu\,.
\ee
Thus we have
\be
dt=H^{4/5}\, d\rho\,,\qquad \td a = H^{4/5}\, a\,,\qquad \td b =
H^{4/5}\, b\,.
\ee

   The original eleven-dimensional Einstein equation is now recast as
the seven-dimensional dilaton equation and Einstein equation.  The
dilaton equation gives
\be
(a^2\, b^4\, (H^4)'\, )' = \ft23 H^4\, \Big({f'}^2 + \fft{2f^2}{a^2} +
\fft{4 f^2\, a^2}{b^4}\Big)\,,
\ee
which can be rewritten as
\be
\fft{6H''}{H} + \fft{18 {H'}^2}{H^2} + \fft{12a'\, H'}{a\, H} + 
\fft{24 b'\, H'}{b\, H} =\fft{{f'}^2}{a^2\, b^4} 
+ \fft{2f^2}{a^4\, b^4} + \fft{4 f^2}{b^8} \,.
\ee
From the Einstein equation we get three separate equations, namely
\crampest
\bea
&&\!\!\!\!\!\!\fft{5a''}{a} + \fft{28 a'\, H'}{a\, H} +
\fft{5{a'}^2}{a^2} 
+\fft{20 a' \, b'}{a\, b} +\fft{16
b'\, H'}{b\, H} + \fft{4 H''}{H} + \fft{12 {H'}^2}{H^2} - \fft5{a^2} -
\fft{5a^2}{b^4}=
\fft{{f'}^2}{4a^2\, b^4} - \fft{2f^2}{a^4\, b^4} + \fft{6
f^2}{b^8}\,,\nn\\
&&\!\!\!\!\!\!\fft{5b''}{b} + \fft{36 b'\, H'}{b\, H} +\fft{15{b'}^2}{b^2} 
+\fft{10 a' \, b'}{a\, b} +\fft{8
a'\, H'}{a\, H} + \fft{4 H''}{H} + \fft{12 {H'}^2}{H^2} - \fft{15}{b^2} 
+\fft{5a^2}{b^4}=
\fft{{f'}^2}{4a^2\, b^4} + \fft{f^2}{2a^4\, b^4} - \fft{4
f^2}{b^8}\,,\nn\\
&&\!\!\!\!\!\!\fft{10a''}{a} + \fft{8 a'\, H'}{a\, H} + 
\fft{20 b''}{b} +\fft{16
b'\, H'}{b\, H} + \fft{24 H''}{H} + \fft{12 {H'}^2}{H^2}=
-\fft{{f'}^2}{a^2\, b^4} + \fft{3f^2}{a^4\, b^4} + \fft{6
f^2}{b^8}\,.
\eea
\uncramp

   This set of equations can be derived from the Lagrangian $L=T-V$,
where
\bea
T&=& -\fft{5\dot a^2}{2a^2} -\fft{20\dot a\,\dot b}{a\, b} -
\fft{15\dot b^2}{b^2} -\fft{20\dot a\,\dot H}{a\, H} -\fft{40\dot b\,
\dot H}{b\, H} - \fft{15 \dot H^2}{H^2} + \fft{5 \dot f^2}{8 a^2\,
b^4}\,,\nn\\
V&=& \ft54 H^8\, \Big(2a^2\, b^4\, (-a^4+ b^4 -6a^2\, b^2) -
(2a^4 + b^4)\, f^2\Big)\,,
\eea
together with the constraint $T+V=0$.  Note that here a dot is a
derivative with respect to the coordinate $\eta$, defined by $d\rho =
a^2\,(b\, H)^4\, d\eta$. The kinetic term $T$ can be expressed as
$T=\ft12 g_{ij} \dot \a_i\, \dot \a_j$, where $a=e^{\a_1}$,
$b=e^{\a_2}$, $H=e^{\a_3}$ and $f=\a_4$, with
\be
g_{ij}= \pmatrix{-5 & -20 & -20 & 0 \cr
                -20 & -30 & -40 & 0 \cr
                -20 & -40 & -30 & 0 \cr 
                  0 &   0 &   0 & \ft{5}{4a^2\, b^4} \cr}
\,.
\ee
Since $g_{ij}$ is field dependent, the system is a non-linear
sigma-model.  Nevertheless, we find that 
the potential $V$ can be expressed in terms
of a superpotential $W$, so that $V=-\ft12\, g^{ij}\, \del_i W\, \del_j
W$, where
\be
W=\ft52 H^4\, (a^2+ b^2) \sqrt{4a^2\, b^4 + f^2}\,.\label{super1}
\ee

\subsection{First-order equations and general solution}

   From the superpotential (\ref{super1}), we can derive
the first-order equations $\dot \a^i = g^{ij}\, \del_j\, W$, which, in
terms of the original radial coordinate $\rho$ of equation 
(\ref{ans1}) become
\bea
&&a'= \fft{6 a^4\, b^4 - 6 a^2\, b^6 + a^2\, f^2 -2 b^2\, f^2}{
    3a\, b^4\, K}\,,\qquad
f' = \fft{2(a^2+b^2)\, f}{K}\,,\nn\\
&&b' = - \fft{12 a^4\, b^4 + 4 a^2\, f^2 + b^2\, f^2}{
    6a^2\, b^3\, K}\,,\qquad
H' =  \fft{(a^2+b^2)\, f^2\, H}{3 a^2\, b^4\, 
K}\,,\label{firstorder1}
\eea
where a prime denotes $d/d\rho$, and we have defined
\be
K\equiv \sqrt{4 a^2\, b^4 + f^2}\,.\label{kdef1}
\ee
 Solutions of these equations will
necessarily satisfy the original second-order equations, implying that
we shall have solutions of the $D=11$ supergravity equations.  One may
note from (\ref{firstorder1}) that
\be
a\, b^2\, H^3\, f = \kappa\,,\label{kappa1}
\ee
where $\kappa$ is a constant of integration.

    In order to solve the first-order equations (\ref{firstorder1}) it
is helpful to define new hatted variables as follows:
\be
a= H^{-1/2}\, \hat a\,,\qquad b= H^{-1/2}\, \hat b\,,\qquad
f=H^{-3/2}\, \hat f\,.
\ee
At the same time, we introduce a new radial variable $\tau$, defined
by $d\tau = H^{1/2}\, d\rho$. The metric ansatz (\ref{ans1}) now
assumes the form
\be
ds_{11}^2 = H^{2}\, dx^\mu\, dx_\mu + H^{-1}\, (d\tau^2 + \hat a^2\,
D\mu^2 + \hat b^2\, d\Omega_4^2)\,.
\ee
The first-order equations (\ref{firstorder1}) are considerably
simplified, becoming
\bea
&&\fft{d\hat a}{d\tau} = \fft{(\hat a^2-\hat b^2)\, \hat K}{2\hat a\, \hat
b^4}\,,\qquad
\fft{d\hat b}{d\tau} = -\fft{\hat K}{2\hat b^3}\,,\nn\\
&&\fft1{\hat f}\, \fft{d\hat f}{d\tau} =  
\fft{(\hat a^2+\hat b^2)\, \hat K}{2\hat a\, \hat b^4}\,,\qquad
\fft{1}{H}\, \fft{dH}{d\tau} = \fft{(\hat a^2 +\hat b^2)\, \hat
f^2}{3 \hat a^2\, \hat b^4\,\hat K}\,,
\eea
where $\hat K \equiv \sqrt{4 \hat a^2\, \hat b^4 + \hat f^2}$.

    A further change of radial coordinate to $r$, defined
by $dr= - \hat K\, d\tau$, puts these
equations in the form
\bea
&&\fft{d\hat a}{dr} = -\fft{(\hat a^2-\hat b^2)}{2\hat a\, \hat
b^4}\,,\qquad \fft{d\hat b}{dr} = \fft{1}{2\hat b^3}\,,\nn\\
&&\fft1{\hat f}\, \fft{d\hat f}{dr} =- 
\fft{(\hat a^2+\hat b^2)}{2\hat a\, \hat b^4}\,,\qquad
\fft{1}{H}\, \fft{dH}{dr} = -\fft{(\hat a^2 +\hat b^2)\, \hat
f^2}{3 \hat a^2\, \hat b^4\,  (4\hat a^2\, \hat b^4 + \hat
f^2)}\,.
\eea
In particular, the equations for $\hat a$ and $\hat b$ have now
decoupled from the rest, and they are in fact nothing but the
first-order equations for the $G_2$ metrics on the $\R^3$ bundle over
$S^4$ (see, for example, \cite{cglphyp}).  

   The system of first-order equations is now completely solvable.
After a final change of radial variable
$r\rightarrow \ft12 r^4$, we find that the general solution for the metric
in the ansatz (\ref{ans1}) is
\be
ds_{11}^2 = H^2\, dx^\mu\, dx_\mu + 2 H^{-7}\, U^{-1}\,
dr^2 + \ft12 r^2\, H^{-1}\, U\,
D\mu^i\, D\mu^i  + r^2\, H^{-1}\, d\Omega_4^2\,,\label{mets2s4sol}
\ee
where
\be
U= 1 -\fft{\ell^4}{r^4}\,,\qquad
H = \Big(1 + \fft{c^2}{2r^{12}\, U^2}\Big)^{1/6}\,,
\ee
and $c$ is a constant. The function $f$ appearing in the 4-form ansatz
(\ref{f4ans}) is given by
\be
f=\fft{c}{r^3\, H^{3/2}\, U^{1/2} }\,.\label{fsol}
\ee
Note that the constant $\kappa$ appearing in (\ref{kappa1}) is precisely
the constant $c$.

\subsection{Properties of the 3-brane solution}

     The general solution contains two non-trivial integration
constants, $\ell$ and $c$.  The constant $\ell$ measures the scale
size of the ``gravitional instanton'' of the background $G_2$
manifold.  The constant $c$, on the other hand, measures the strength
of the the 4-form field $F_\4$.  Asymptotically at large distance, the
gravitational instanton contributions, going as $\ell^4/r^4$, dominate
in comparison to the $F_\4$ contribution, going as $c^2/r^{12}$.

       When the constant $c$ is set to zero, the 4-form field is
turned off, and consequently the configuration reduces to the vacuum
solution of a direct product $M_4\times {\cal M}_7$ of Minkowski
4-spacetime and the seven-dimensional smooth manifold with $G_2$
holonomy.  When $c$ is non-vanishing, the solution describes a 3-brane
in $D=11$, with a four-dimensional Poincar\'e symmetry in its world
volume.  Asymptotically, the solution approaches $M_4\times {\cal
M}_7$.  If the parameter $\ell$ is taken to be zero, then the smooth
7-manifold ${\cal M}_7$ has a singular limit to the cone over the
$S^2$ bundle over $S^4$ (or $\CP^2$).  At small distance, near $r=\ell$, the
metric becomes singular, with the limiting form
\bea
\ell=0:&& ds_{11}^2 =\rho^{-\ft12}\, dx^\mu\wedge dx_\mu +
\rho^{\ft12}\, (\ft12 D\mu^i\, D\mu^i + d\Omega_4^2)+ d\rho^2\,,\nn\\
\ell\ne 0:&& ds_{11}^2 = \rho^{-\ft25}\, dx^\mu\wedge dx_\mu +
\ft 12\rho^{\ft15}\, D\mu^i\, D\mu^i + \rho^{\ft45}\,
d\Omega_4^2+ d\rho^2
\eea
as the proper distance $\rho$ tends to zero.   
We can also consider the the possibility of
sending $\ell^4 \longrightarrow -\ell^4$.  In this case, the metric 
behaviour at small proper distance (\ie near $r=0$) becomes
\be
ds_{11}^2 = \rho^{-1/4}\, (dx^\mu\, dx_\mu + \ft12 D\mu^i\, D\mu^i)
+ \rho^{1/2}\, d\Omega_4 + d\rho^2\,.
\ee

       The 4-form flux is given by
\be
Q=2f\, \int_{S^4} \Omega_\4\,,
\ee
and from (\ref{fsol}) this can be seen to vanish when $r$ is sent to
infinity.  Thus our M3-brane configurations do not carry any conserved
charge, and might be described as ``3-branes without 3-branes.''
The solution should really be thought of as a gravitional monopole
involving the supergravity multiplet.  The two constants $\ell$ and
$c$ are both continuous parameters.

     What is perhaps the most interesting feature of the solution is
that it is massless.  The leading-order $r$ dependence in the function
$H$ at large $r$ is $c^2/r^{12}$, whilst a mass term for a
7-dimensional transverse space would have a leading-order
$r$-dependence of the form $m/r^{5}$.  The existence of the naked
singularity may be related to the fact that the solution is massless.
In fact all the previously known massless $p$-brane solutions in
supergravity contain naked singularities
\cite{beh,kallind,cveticyoum1,cveticyoum,dyonicstring}.  A repulson
mechanism was proposed in string theory \cite{repul} to resolve such a
naked singularity in the massless dyonic string \cite{dyonicstring}.
A further observation is there appears not to exist a natural and
non-trivial decoupling limit.  This may also be a consequence of the
masslessness.  For example, a massive dyonic string has a decoupling
limit, but this ceases to exist when the charges are tuned for
masslessness.

   Unlike the dyonic string, where the masslessness is achieved by
making an adjustment of integration constants, in our new M3-brane
solution there {\it is} no mass integration constant.  From the point
of view of supersymmetry, the masslessness is consistent with the
absence of a non-vanishing conserved 4-form charge.  However, we shall
defer a more detailed investigation of the supersymmetry of this
solution until section 4.

\section{M3-brane in background of $\R^4$ bundle over $S^3$}

   In this section we construct an analogous 3-brane solution in the
background of the third manifold of $G_2$ holonomy, whose topology is
$\R^4\times S^3$.  As we shall see, the configuration is again a 
``no-braner,'' which carries no conserved brane charge.

\subsection{The ansatz}

   In this case, we consider the eleven-dimensional metric ansatz
\be
ds_{11}^2= H^2\, dx^\mu\, dx_\mu + d\rho^2 + a^2\, \nu_i^2 + b^2\,
\Sigma_i^2\,\label{metans2}
\ee
where $\nu_i\equiv\sigma_i -\ft12\Sigma_i$, and $\sigma_i$ and
$\Sigma_i$ are two sets of left-invariant on two independent $SU(2)$
group manifolds.  The level surfaces $r=$ constant are therefore
an $S^3$ bundle over $S^3$.  Since the bundle is a trivial one, the
level surfaces are topologically $S^3\times S^3$.  There is
a vacuum solution which is a complete Ricci-flat manifold, namely the direct
product of four-dimensional Minkowski spacetime 
and the known seven-dimensional manifold $\R^4\times S^3$ with
$G_2$ holomony \cite{brysal,gibpagpop}.  

    Again we should now like to turn on the 4-form field
strength, in order to introduce a 3-brane configuration in this background.
The ansatz for the 4-form, respecting the symmetries of the vacuum,
can be written as \cite{clptrans}
\be
F_\4 = f_1\, \nu_i\wedge \nu_j\wedge \Sigma_i\wedge \Sigma_j +
         f_2\, d\rho\wedge \nu_1\wedge \nu_2\wedge \nu_3 +
         \ft12 f_3\, \epsilon_{ijk}\, d\rho\wedge \nu_i\wedge
          \Sigma_j\wedge\Sigma_k\,.\label{f4s3s3}
\ee
The Bianchi identity $dF_\4=0$ gives
\be
f_1' - \ft18 f_2 + \ft12 f_3=0\,,
\ee
and the field equation $d{\hat * F_\4}=0$ gives
\be
\fft{2 f_1\, H^4}{a\, b} +  (f_3\, a\, b^{-1}\, H^4)' =0\,,\qquad
-\fft{3 f_1\, H^4}{a\, b} + (2 f_2\, b^3\, a^{-3}\, H^4)' =0\,.
\label{f123eq}
\ee
It is again convenient to derive the conditions implied by the
eleven-dimensional Einstein equation in terms of a dimensional
reduction to $D=7$.  The dilaton equation gives
\be
\fft{H''}{H} + \fft{3 {H'}^2}{H^2} + \fft{3 a'\, H'}{a\, H} 
   + \fft{3 b'\, H'}{b\, H} - \fft{2 f_1^2}{a^4\, b^4} - 
\fft{f_2^2}{6 a^6} - \fft{f_3^2}{2 a^2\, b^4}=0\,,
\ee
and finally, the seven-dimensional Einstein equation gives
\crampest
\bea
\!\!\!&&\!\!\!\fft{5 b''}{b} + \fft{4 H''}{H} + 
\fft{15 a'\, b'}{a\, b} + \fft{10
{b'}^2}{b^2} + \fft{12 a'\, H'}{a\, H} + \fft{32 b'\, H'}{b\, H} 
+ \fft{12 {H'}^2}{H^2} -\fft{5}{2b^2} +\fft{5 a^2}{16 b^4} +
\fft{2f_1^2}{a^4\, b^4} - \fft{3 f_2^2}{2 a^6} + \fft{f_3^2}{2 a^2\,
b^4} =0\,,\nn\\
\!\!\!&&\!\!\!\fft{5 a''}{a} + \fft{4 H''}{H} + \fft{15 a'\, b'}{a\, b} 
+ \fft{10
{a'}^2}{a^2} + \fft{32 a'\, H'}{a\, H} + \fft{12 b'\, H'}{b\, H} 
+ \fft{12 {H'}^2}{H^2} -\fft{5}{2a^2} -\fft{5 a^2}{32 b^4} +
\fft{2f_1^2}{a^4\, b^4} +\fft{f_2^2}{a^6} - \fft{2f_3^2}{ a^2\,
b^4} =0\,,\nn\\
\!\!\!&&\!\!\!\fft{15 a''}{a} + \fft{15 b''}{b} + \fft{24 H''}{H} 
+ \fft{12 a'\,
H'}{a\, H} + \fft{12 b'\, H'}{b\, H} + \fft{12 {H'}^2}{H^2} 
-\fft{18 f_1^2}{a^4\, b^4} + \fft{f_2^2}{a^6} +\fft{3 f_3^2}{a^2\,
b^4}=0\,.
\eea
\uncramp

\subsection{First order equations and general solution}

    From (\ref{f123eq}), we can solve for $f_1$ and $f_2$,
\be
f_1= -\ft12 a\, b\, H^{-4} (f_3\, a\, b^{-1}\, H^4)'\,,\qquad
f_2 = -\ft34 a^3\, b^{-3}\, H^{-4}\, ( \lambda + f_3\, a\, 
b^{-1}\, H^4)
\,,\label{f1f2sol}
\ee
where $\lambda$ is a constant of integration.
The remaining equations for $a$, $b$, $H$ and $f_3$ can then be
obtained from the Lagrangian $L=T-V$, together with the constraint
$T+V=0$, where $T=\ft12 g_{ij}\,\dot \a^i\, \dot \a^j$ with
$\a^i=(\log a, \log b, \log H, f_3)$, and a dot denotes a derivative with
respect to $\eta$ defined by $dt=a^3\,b^3\, H^4\, d\eta$.  We have
\be
g_{ij} = \pmatrix{-60-\fft{15f_3^2}{b^4} & 
                  -90 + \fft{15f_3^2}{b^4} &
                  -120 - \fft{60f_3^2}{b^4} &
                  -\fft{15f_3}{b^4} \cr
            -90 + \fft{15f_3^2}{b^4} &
            -60 - \fft{15f_3^2}{b^4} &
            -120 + \fft{60f_3^2}{b^4} &
               \fft{15f_3}{b^4} \cr
                  -120 - \fft{60f_3^2}{b^4} &
                 -120 + \fft{60f_3^2}{b^4} & 
                 -120 - \fft{240f_3^2}{b^4} &
                    -\fft{60f_3}{b^4} \cr
                -\fft{15f_3}{b^4} &
                \fft{15f_3}{b^4} &
                -\fft{60f_3}{b^4} & -\fft{15}{b^4}}\,.
\ee
The potential $V$ is given by
\be
V=\ft{15}{32} H^8\, a^4\, (-a^4\, b^2 + 16a^2\, b^4 + 16 b^6 +
3a^4\, b^{-2}\, f_3^2 + 16b^2\, f_3^2) + \fft{45}{32}\lambda\,
a^6\, (\lambda -2a\, b^{-1}\, H^4\, f_3)\,.
\ee

As in the previous case, the kinetic term $T$ is of the form of a
non-linear sigma model, with a fairly complicated field-dependent
$g_{ij}$.  The inverse is relatively simpler, given by
\be
g^{ij} = \pmatrix{\ft1{45} & -\ft1{90} & -\ft1{90} & \ft1{90}f_3 \cr
                 -\ft1{90} &\ft1{45} & -\ft1{90} & \ft7{90}f_3 \cr
              -\ft1{90} & -\ft1{90} & \ft1{72} & -\ft1{18}f_3\cr
          -\ft1{90}f_3 & \ft7{90}f_3 & -\ft1{18}f_3 &
           -\ft3{45}b^4 + \ft{13}{45} f_3^2}
\ee

If the integration constant $\lambda$ is taken to be zero,\footnote{If
the integration constant $\lambda$ were non-vanishing, which would
correspond to a configuration including M5-branes wrapped on 3-cycles
in $S^3\times S^3$, it is not clear how one would solve the
second-order equations.  Similar remarks apply to the previous case in
section 2 also.  We chose to omit an analogous constant of
integration, in the discussion above (\ref{f4ans}), in order to obtain
a formulation of the second-order equations in terms of a
superpotential and hence a gradient flow.  Had we retained the
constant of integration, which would give a non-vanishing flux
$\lambda\, \Omega_\4$ for $F_\4$ corresponding to M5-branes wrapping
on 2-cycles in $\CP^3$ or $SU(3)/(U(1)\times U(1))$, it is again not
clear how one would solve the second-order equations.  We thank
S.S. Gubser for raising this question about our
procedure for obtaining gradient flows.} we find that the potential
can be expressed in terms of a superpotential $W$, {\it i.e.}\
$V=-\ft12g^{ij}\, \del_i W\, \del_j W$, with
\be
W=\ft{15}4 H^4\, a^2\, b^{-1}\, (a^2 + 4b^2)\, \sqrt{b^4 -f_3^2}
\,.
\ee
From this we can obtain first-order equations, given by
\bea
&&a' = \fft{a^2\, b^4 - 4b^6 -2a^2\, f_3^2}{8b^4\, K}\,,\qquad
b' = \fft{-2a^2\, b^4 + (a^2 -4b^2)\, f_3^2}{8a\, b^3\, K}
\,,\nn\\
&&H' = \fft{H\,(a^2 + 4b^2)\, f_3^2} {8a\, b^4\,K}\,,\qquad
f_3' = \fft{f_3\Big((12b^2-a^2)b^4-f_3^2(a^2+20b^2)\Big)}{8
             a\, b^4\, K}\,,\label{abhf}
\eea
where $K\equiv \sqrt{b^4-f_3^2}$ and a prime denotes a derivative with
respect to the original $\rho$ coordinate appearing in (\ref{metans2}).
Note that these first-order equations again imply an algebraic
relation among the functions, analogous to (\ref{kappa1}).  This time,
we have
\be
a^3\, b\, H^6\, f_3 = \kappa\,.\label{kappa2}
\ee

    The equations can be solved by defining new quantities $\hat a
\equiv H\, a$,  $\hat b\equiv H\, b$ and $\hat f_3 \equiv H^2\, f_3$.
After manipulations analogous to those in section 2, we arrive at the
general solution
\be
ds_{11}^2 = H^2\, dx^\mu\, dx_\mu + 12 H^4\, U^{-1}\,  dr^2
           + \ft43 r^2\, H^{-2}\, U\, \nu_i^2 + r^2\, H^{-2}\,
\Sigma_i^2\,,\label{mets3s3sol}
\ee
where
\be
U\equiv 1-\fft{\ell^3}{r^3} \,,\qquad 
H = \Big( 1 -\fft{c^2}{r^{12}\, U^3}\Big)^{-1/6}\,.
\ee
The function $f_3$ is given by
\be
f_3= \fft{c}{r^4\, H^2\, U^{3/2}}\,.\label{f3s3s3sol}
\ee
Note that the constant $\kappa$ in (\ref{kappa2}) is related to $c$ by
$\kappa=8c/(3\sqrt3)$.

    Thus we see that the general solution has two non-trivial
integration constants, $\ell$ and $c$.  The constant $\ell$ measures the
scale size of the the gravitional instanton of the $G_2$ manifold,
whilst the constant $c$ measures the contribution from the 4-form field
strength.  Again, the solution is massless and carries no charge, and it can
be thought of as a gravitional monopole involving the supergravity
multiplet fields.  Asymptotically, the solution a becomes a product of
four-dimensional Minkowski spacetime and the $G_2$ manifold with
$\R^4\times S^3$ topology, since the contribution to the metric is
dominated at large $r$ by the instanton contribution $\ell^4/r^4$, in
comparison to the $F_\4$ contribution which is of order $c^2/r^{12}$.
At small distance, the solution has a naked singularity.

\section{Supersymmetry of the M3-branes}

   Since the configurations that we have obtained in the
previous two sections arise as the solutions of first-order systems of
equations, it is natural to expect that they should be
supersymmetric.  In other words, one would expect that the first-order
equations would have the interpretation of being precisely the
integrability conditions for supersymmetry.  However, since they were
not obtained by explicitly requiring supersymmetry, but rather by
finding a superpotential for the Lagrangian formulation of the
original bosonic supergravity equations of motion, the question
of supersymmetry remains to be investigated.

\subsection{Solutions in the $\R^4$ bundle over $S^3$ background}

   First, we shall study the supersymmetry for the solutions obtained in
section 3, where the background metric in the transverse space is the
$\R^4$ bundle over $S^3$.  It is a straightforward matter to calculate
the spin connection for the metric (\ref{metans2}) directly in eleven
dimensions, and then to substitute this and the field strength ansatz
(\ref{f4s3s3}) directly into the gravitino transformation rule
\be
\delta\hat \psi_A = D_A\, \hat\ep - \fft{1}{288}\, F_{BCDE}\,
\hat \Gamma_A{}^{BCDE}\, \hat\ep + 
   \fft{1}{36} \, F_{ABCD}\, \hat\Gamma^{BCD}\,
\hat\ep\,.\label{susytrans}
\ee
We make a standard $4+7$ 
decomposition of the Dirac matrices, as follows:
\be
\hat\Gamma_\mu = \gamma_\mu\times\oneone\,,\qquad
\hat\Gamma_a= \gamma_5\times \Gamma_a\,.\label{gammadecomp}
\ee
Substituting into (\ref{susytrans}), and examining first the
world-volume directions $\mu$, we find that a Killing spinor of the form
$\hat\ep=\ep\times \eta$ must satisfy $\gamma_5\, \ep=\pm\ep$.  For
the case of $\gamma_5\, \ep=+\ep$ we find
\be
\eta= g\, (b^2+f_3)^{1/2} \, \eta_1 + g\, (b^2-f_3)^{1/2}\, \eta_2\,,
\label{wvdir}
\ee
where $\eta_1$ and $\eta_2$ are constant spinors in the transverse
7-space, satisfying the projection conditions
\be
(\Gamma_1 +\im\,
\Gamma_4)\, \eta_2=0\,,\quad \Gamma_{23}\, \eta_1 =-\Gamma_{04}\,
\eta_2\,,\quad (\Gamma_{26}-\Gamma_{35})\, \eta_1 =-2\im\,
\Gamma_{04}\, \eta_2\,.
\ee
Note that these conditions uniquely determine $\eta_1$ and $\eta_2$,
up to an overall scale.  For the case where $\gamma_5\, \ep=-\ep$, the
associated spinor $\eta$ is given again by (\ref{wvdir}), but with
$f_3$ replaced by $-f_3$.
The dependence of the overall function $g$ on the coordinates of the
transverse space is undetermined by the $\mu$ components of
$\delta\hat \psi_A=0$.

   The components of $\delta\hat\psi_A=0$ lying in the directions
$A=a$ of the transverse space will now determine the dependence of $g$
on the transverse coordinates.  It is easiest first to examine the
radial direction (\ie the ``0'' direction), which determines the
radial dependence of the Killing spinor.  Then, by looking at the
remaining transverse directions, we find that the Killing spinor has
no dependence on the angular coordinates of the two 3-spheres.  The
conclusion is that the function $g$ in (\ref{wvdir}) is given by
\be
g = b^{-1}\, H^{1/2}\,.
\ee
Thus the first-order equations (\ref{f1f2sol}) (where $f_1$ and $f_2$
are given by (\ref{abhf}) with $\lambda=0$) are precisely the
integrability conditions for the existence of a spinor $\hat\ep$
satisfying $\delta\hat\psi_A=0$ in (\ref{susytrans}).  Since we have 
2 solutions
(corresponding to two spinors $\ep$ in the M3-brane world-volume) for
each of the cases $\gamma_5\, \ep=\pm\ep$, the general Killing spinor
has four real solutions, corresponding to $N=1$ supersymmetry on the
world-volume of the M3-brane.  Of course if the constant $c$ is set to
zero, so that the 4-form is turned off, these Killing spinors reduce
to the usual ones in the product of four-dimensional Minkowski
spacetime and the Ricci-flat metric of $G_2$ holonomy.

\subsection{Solutions in the $\R^3$ bundle over $S^4$ background}

   Here, we repeat the analysis of the supersymmetry transformations
in the case of the $\R^3$ bundle over $S^4$ background, for which the
M3-brane solution was constructed in section 2.

    Again we begin by considering the components of
$\delta\hat\psi_A=0$ lying in the world-volume of the 3-brane.  From
$\delta\hat\psi_\mu=0$ we deduce that a Killing spinor of the form
$\hat\ep=\ep\times \eta$ will be given by $\gamma_5\, \ep=\pm\ep$,  
and for the case $\gamma_5\ \ep=+\ep$ we have 
\be
\eta = g\, P\, \Big[ (K-f)^{1/2} \, \eta_1 + (K+f)^{1/2}\, \eta_2\Big]\,,
\label{secondsol}
\ee
where $K$ is given by (\ref{kdef1}), $P$ is given by
\bea
P&\equiv& 
 -\fft{\sin\ft12\theta}{2 a\, b^2}\, \Big[ (K\cos\ft12\phi -\im\, f\,
\sin\ft12\phi) \, \Gamma_{01} +  (K\cos\ft12\phi +\im\, f\,
\sin\ft12\phi) \, \Gamma_{02}\Big]\nn\\
&&+\cos\ft12\theta\, (\cos\ft12\phi + \sin\ft12 \phi\, \Gamma_{12}) 
\,,\label{Psol}
\eea
and the constant 8-component spinors $\eta_1$ and $\eta_2$ are
uniquely specified (up to scale)  by the projections
\bea
&&(\Gamma_0-\Gamma_{3456})\, \eta_2=0\,,\qquad 
(\Gamma_{34} + \Gamma_{56} -2 \Gamma_{012})\,
\eta_2=0\,,\nn\\
&&4 \eta_1 = (\Gamma_{135} -\Gamma_{146} +\Gamma_{236}
+\Gamma_{245})\, \eta_2\,.
\eea
Here the explicit indices 1 and 2 on the Dirac matrices refer to the
two directions on the $S^2$ fibres, as in (\ref{s2viel}), whilst the indices 
3, 4, 5 and 6 refer to the directions in the $S^4$ base.  The index 0
refers to the radial direction.  For the case when $\gamma_5\,
\ep=-\ep$, the associated spinor $\eta$ in the transverse space will
be given again by (\ref{secondsol}), but now with $f$ sent to $-f$ in
(\ref{secondsol}) and in (\ref{Psol}).

   The dependence of the overall prefactor $g$ on the coordinates of the
transverse space is not determined by $\delta\hat\psi_\mu=0$ in the
world-volume directions.  (We have, however, made a convenient choice
of $\theta$ and $\phi$ dependent overall factors, in anticipation of
subsequent results.)

   From the radial component $\delta\hat\psi_0=0$ in the transverse
space, we can again determine the radial dependence of the 
function $g$, finding
\be
g =\fft{\hat g}{H\, b\, a^{1/2}}\,,
\ee
where $\hat g$ depends only on the angular coordinates of the
transverse space.  Examination of $\delta\hat\psi_A=0$ in the $S^2$
directions then implies that $\hat g$ is independent of these two
coordinates.  Finally, the components in the $S^4$ directions
determine that $\hat g$ has the dependence associated with the singlet
fermion zero-mode in the Yang-Mills instanton background (as in 
\cite{cglpnew}). 

    We have seen that again, the first-order system of equations for
this 3-brane in the background of the $\R^3$ bundle over $S^4$ have
turned out to be precisely the integrability conditions for the
existence of a Killing spinor.  There are in total four real
solutions, implying $N=1$ supersymmetry on the world-volume of the
M3-brane.  As in the previous example, if the field strength in the
solution is taken to zero, by setting the constant $c=0$, the Killing
spinor reduces to the standard one in the vacuum of four-dimensional
Minkowski spacetime times the Ricci-flat metric on the $\R^3$ bundle
over $S^4$.

\section{Dual formulations and phase transitions}

   In a standard massive BPS $p$-brane solution, the charge $Q$ arises
as a constant prefactor in the field strength supporting the solution,
and $Q$ also appears linearly in the harmonic function $H$ in the
$p$-brane metric.  By contrast, in our massless M3-brane solutions the
analogous constant $c$ that arises as the prefactor in the expressions
for the 4-form field appears quadratically in the metrics
(\ref{mets2s4sol}) and (\ref{mets3s3sol}).  This means that the
metrics would continue to be real if we were to send $c\longrightarrow
\im\, c$.  The same would also be true of the reduced metrics in $D=7$
that formed the starting-points of our derivations in sections 2 and 3.

    Of course sending $c\longrightarrow \im\, c$ would imply that the
4-form field strength would become imaginary.  However, it should be
recalled that our original 7-dimensional starting point was in the
Euclidean-signatured theory obtained by dimensional reduction on the
world-volume of the M3-brane.  In Euclidean signature, if the 4-form
field strength $F_\4$ is dualised to a 3-form $F_\3$, then 
its kinetic term in the $D=7$ Lagrangian will undergo the replacement 
\be
-\ft1{48}\, e^{\sqrt{8/5}\, \phi}\, F_\4^2 \longrightarrow 
+\ft1{12}\, e^{-\sqrt{8/5}\, \phi}\, F_\3^2 \,.\label{dualsign}
\ee
This change of sign of the kinetic term, which is generic to all
dualisations in Euclidean signature, indicates that we could achieve
the same effect as sending $c\longrightarrow \im\, c$ by instead 
using a {\it real} 3-form field in $D=7$, but with the canonical 
$-\ft1{12}\, e^{-\sqrt{8/5}\, \phi}\, F_\3^2$ kinetic term instead of
the sign-reversed one  in (\ref{dualsign}) that arose by dualising the
4-form.

    The upshot of the above discussion is that we can obtain real solutions
in $D=7$ that are just like those in sections 2 and 3, but for the
opposite sign of $c^2$.  These will be solutions of the equations
coming from the seven-dimensional Lagrangian
\be
e^{-1}\, \wtd{\cal L}_7 = R - \ft12 (\del\phi)^2 - \ft1{12}\, 
e^{-\sqrt{\fft85}\, \phi}\, F_\3^2\,.\label{d7lag2}
\ee
The expression for $F_\3$ for each solution will be given by $F_\3 =
-\im\, {*F_\4}$, where $F_\4$ is the corresponding expression given in
section 2 or 3.  The $\im$ factor in this relation between $F_\3$ and
$F_\4$ is precisely removed by the $\im$ factor that we acquire upon
sending $c\longrightarrow \im\, c$.

   A difference now arises when we consider the higher-dimensional
origin of the seven-dimensional Lagrangian.  We viewed (\ref{d7lag1})
in sections 2 and 3 as coming from the Kaluza-Klein reduction of $D=11$
supergravity on the world-volume of the M3-brane.  Instead, we should
now view (\ref{d7lag2}) as coming from the Kaluza-Klein reduction of 
type IIA, type IIB or type I supergravity on the world-volume of a
2-brane.  In other words, we obtain the 3-form in $D=7$ as the direct
world-volume reduction of a 3-form in $D=10$.  Accordingly, we can
then lift the $D=7$ solutions of sections 2 and 3, after sending
$c\longrightarrow \im\, c$, to real solutions of ten-dimensional
supergravity.  Thus there is a phase transition from one type of brane
to another, when we change the modulus parameter $c$ of the solution
from real to imaginary.

   For the case of the $S^2$ bundle over $S^4$ in section 2, we find
that the corresponding massless 2-brane solution in $D=10$ is given by
\be
ds_{10}^2 = H^{-3/2}\, dx^\mu\, dx_\mu + 2 H^{-9/2}\, U^{-1}\,
dr^2 + \ft12 r^2\, H^{3/2}\, U\,
D\mu^i\, D\mu^i  + r^2\, H^{3/2}\, d\Omega_4^2 \,,\label{mets2s4sol10}
\ee
where
\be
U= 1 -\fft{\ell^4}{r^4}\,,\qquad
H = \Big(1 - \fft{c^2}{2r^{12}\, U^2}\Big)^{1/6}\,.\label{uh1}
\ee
For the $S^3$ bundle over $S^3$ of section 3, the corresponding
massless 2-brane solution in $D=10$ is given by
\be
ds_{10}^2 = H^{-3/2}\, dx^\mu\, dx_\mu + 12 H^{13/2}\, U^{-1}\,  dr^2
           + \ft43 r^2\, H^{1/2}\, U\, \nu_i^2 + r^2\,H^{1/2}\, 
\Sigma_i^2\,,\label{mets3s3sol10}
\ee
where
\be
U\equiv 1-\fft{\ell^3}{r^3} \,,\qquad 
H = \Big( 1 +\fft{c^2}{r^{12}\, U^3}\Big)^{-1/6}\,.\label{uh2}
\ee
We have written the solutions that come from reducing the NS-NS 3-form 
of the ten-dimensional supergravity.  Of course in the case of type
IIB we could instead use the R-R 3-form, in which case the lifted
solutions in $D=10$ would simply be the S-duals of those we have just
presented. 

   One can also, of course, further lift the above configurations, if
viewed as solutions of type IIA supergravity, to $D=11$.   For the
case corresponding to the $S^2$ bundle over $S^4$ we then find
\be
ds_{11}^2 = H^{-2}\, dx^\mu\, dx_\mu + H^4\, dz^2 + 2 H^{-5}\, U^{-1}\,
dr^2 + \ft12 r^2\, H\, U\,
D\mu^i\, D\mu^i  + r^2\, H\, d\Omega_4^2 \,,\label{mets2s4sol11}
\ee
where $z$ is the eleventh coordinate, and $U$ and $H$ are again given
by (\ref{uh1}).  For the case corresponding to the $S^3$ bundle over
$S^3$, we find
\be
ds_{11}^2 = H^{-2}\, dx^\mu\, dx_\mu + H^4\, dz^2 + 
12 H^{6}\, U^{-1}\,  dr^2
           + \ft43 r^2\, U\, \nu_i^2 + r^2\, 
\Sigma_i^2\,,\label{mets3s3sol11}
\ee
where $U$ and $H$ are given by (\ref{uh2}).

\section{$D=7$ 3-brane and $S^2$-wrapped M5-brane}

      It is of interest also to study more general 3-brane
configurations in M-theory. Another natural candidate is an M5-brane
wrapped around a supersymmetric 2-cycle.  M5-branes wrapped on
supersymmetric cycles have been discussed previously
\cite{maldnunez1,gaukimwald}.  Typically, they admit solutions of the
form AdS$_d\times H_{7-d}$, where $H_n$ denotes the $n$-dimensional
hyperbolic space.  In this section, we shall consider an M5-brane
wrapped around a 2-sphere.  The solution can be obtained by looking
first at gauged supergravity in $D=7$.

\subsection{$D=7$ AdS$_7$ 3-brane}

    Consider the $D=7$, $N=2$ gauged supergravity, whose bosonic Lagrangian
is
\bea 
{\cal L}_7 &=& R{*\oneone} -\ft12 {*d\phi}\wedge d\phi -
U\,{*\oneone}-\ft12 e^{-\ft{4}{\sqrt{10}}\phi} {*F_\4\wedge F_\4}\nn\\
&&-\ft12 e^{\ft{2}{\sqrt{10}}\phi} {*F_\2}\wedge F_\2 +\ft12
F_\2^i\wedge F_\2^i \wedge A_\3 -\ft1{2\sqrt2}g\, F_\4\wedge\
A_\3\,.\label{d7lag} 
\eea
where $F_\4=dA_\3$ and $U$ is the scalar potential in the $D=7$ gauged
supergravity,
\be
U= g^2\,(\ft14 e^{\fft{8}{\sqrt{10}}\, \phi} - 
   2e^{\fft{3}{\sqrt{10}}\, \phi} -
   2 e^{-\fft{2}{\sqrt{10}}\, \phi})\,.\label{scalarpot}
\ee
In addition, the 4-form satisfies the first-order odd-dimensional self-duality
equation
\be
e^{-\ft{4}{\sqrt{10}}\phi}\, {*F_\4} =-\ft1{\sqrt2}g\, A_\3 + \ft12
\omega_\3\,.
\ee
Here, we have $\omega_\3\equiv A_\1^i\wedge F_\2^i -\ft16 g\,
\epsilon_{ijk} A_\1^{i} \wedge A_\1^j\wedge A_\1^k$.  Domain wall and
AdS$_7$ black hole solutions in this theory have been constructed
\cite{ten,minliu,dist}, which can be viewed after lifting back to 
M-theory as distributed or rotating M5-branes respectively.

     Here we consider a 3-brane configuration, which is supported by
one component of the $SU(2)$ Yang-Mills gauge fields.  We take the ansatz
to be
\bea
ds_7^2 &=& e^{2A}\, dx^\mu\wedge dx_\mu + e^{2B}\, (dr^2 + d\Omega_2^2)
\,,\\
F_\2^3 &=& \lambda\, \Omega_2\,,\qquad F_\4=0\,.
\eea
The resulting equations of motion can be derived from the
Lagrangian $L=T-V$, where
\bea
T &=& -12 \dot A^2 -16 \dot A\, \dot B -2 \dot B^2 +\ft12
\dot \phi^2\,,\nn\\
V&=&e^{8A+2B}(2 - e^{2B}\, U - \ft12\lambda^2\, 
e^{-2B + \fft{2}{\sqrt{10}}\, \phi})\,,\label{abelianlag}
\eea
together with the constraint $T+V=0$.  Here the dot denotes a derivative with
respect to $\eta$, defined by $d\eta =e^{8A+2B}\, dr$.   

    We find that $V$ can be derived from a superpotential $W$,
provided that $g\,\lambda=1$. It is given by
\bea
W=2\sqrt2\, g\,e^{4A+2B-\ft1{\sqrt{10}}\phi} +
\sqrt2\, g^{-1}\, e^{4A +\ft1{\sqrt{10}}\phi} +
\ft1{\sqrt2}\, g\,e^{4A+2B+\ft4{\sqrt{10}}\phi}\,.
\eea
The associated first-order equations, after setting $g=1$ without loss
of generality, are given by
\be
\fft{a'}{a} = -b^2\, f^4 - \fft{4b^2}{f} + 2f\,,\quad
\fft{b'}{b} = - b^2\,f^4 - \fft{4b^2}{f} -8f\,,\quad
\fft{f'}{f} = 4b^2\, f^4 - \fft{4b^2}{f} + 2f\,,\label{abfeqs}
\ee
where $a=e^{A}$, $b=e^{B}$, $f=e^{\ft1{\sqrt{10}}\phi}$, and a prime here
denotes a derivative with respect to $\rho$, which is defined by
$d\rho = \fft1{10\sqrt2} e^{-B}\, dr$.   

     It is not clear how to solve these first-order equations
analytically, but the general behaviour of the solutions to the
gradient flow can
nevertheless be analysed in terms of a phase-plane diagram.   
From (\ref{abfeqs}), we can plot the 2-dimensional vector $(b',
f')$, and it shows that the solution flows from $(b\rightarrow
\infty, f\rightarrow 1)$ to $(b\rightarrow 0, f\rightarrow
\infty)$. (See figure 1. Note that $f$ is always non-negative.)

\vskip 1cm
\psfig{figure=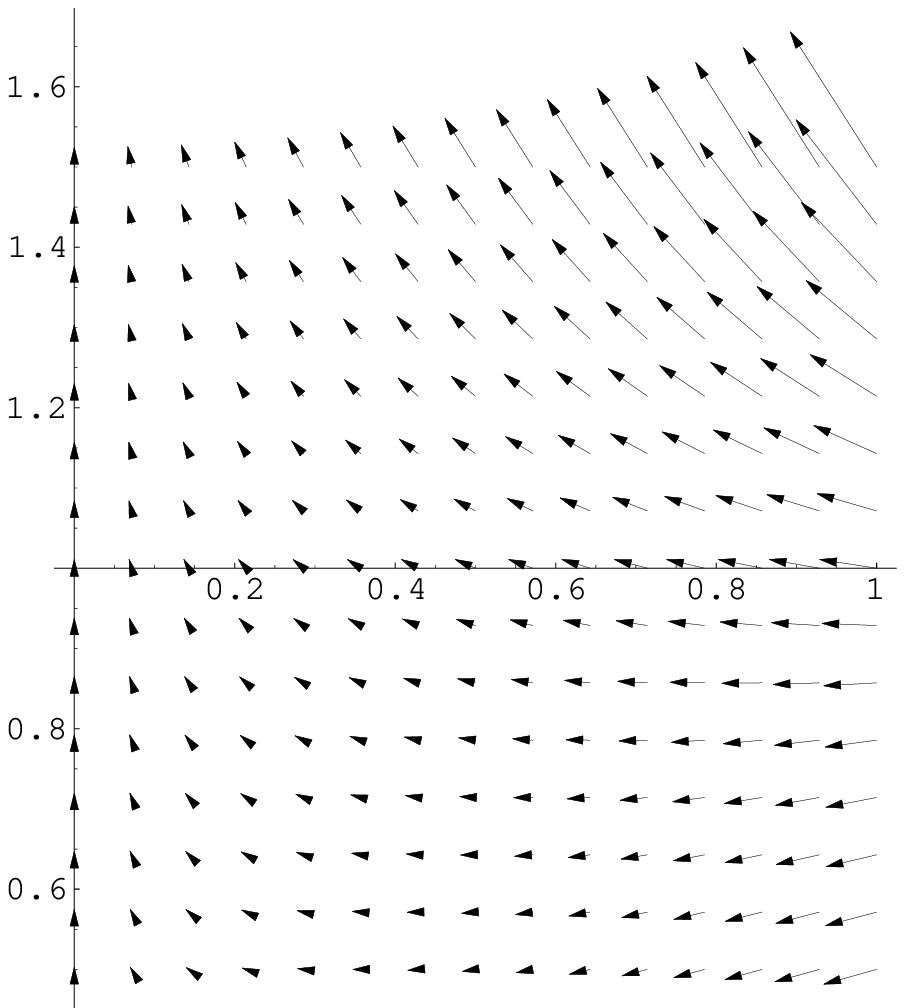}
\vskip 1 cm
\centerline{Figure 1. The flow defined by the 2-vector $(b',f')$. 
The abscissa is $b$ and the ordinate is $f$.}
\vskip 1 cm

    It suffices to analyse the solution in the regions 
$(b\rightarrow \infty, f\rightarrow 1)$ and $(b\rightarrow 0,
f\rightarrow \infty)$.   Consider first the behaviour when
$f\rightarrow 1$.  In this case, the approximate form of the solution is
\bea
a^2 &\sim& \fft{1}{10\rho}+\fft65 + \fft{181\rho}{30}+ \cdots\,,\qquad
b^2 \sim \fft{1}{10\rho} -\fft45 + \fft{11\rho}{15} + \cdots\,,\nn\\
f   &\sim & 1-2\rho + \rho^2(\ft{638}{21} + 40 \log\rho) 
-1120 \rho^3 \log\rho + \cdots\,,
\eea
up to the first few orders in $\rho$.
It is easy to see that $r\sim \sqrt{\rho}\rightarrow 0$, and so
the metric approaches
\be
ds_7^2 \sim \fft1{r^2}\, (dx^\mu\, dx_\mu + d\Omega_2^2+ dr^2)\,. 
\ee
Since $r$ tends to zero here, this describes the large-distance
asymptotic region.

   Now consider the behaviour when $f\rightarrow \infty$, 
with $b$ approaching zero.  In this case, we have
\be
a^2 \sim (1-\rho)^{1/7}\sim b^2\,,\qquad f\sim \ft12 (1-\rho)^{-2/7}\,.
\ee
with $\rho$ tending to 1 from below.  The metric then has the form
\be ds_7^2 = (r-r_0)^{2/15}\, (dx^\mu\, dx_\mu + d\Omega_2^2 +dr^2)
\,.  \ee Since $r\longrightarrow r_0$ in this case, it clearly
corresponds to the region at small proper distance.  The solution at
$r=0$ would be singular, but it is also a horizon

   We can dimensionally reduce the solution on $d\Omega_2^2$, to
obtain a domain wall in $D=5$, of the form
\be 
ds_5^2 = e^{2C} (dx^\mu\,dx_\mu + dr^2)\,.  
\ee 
The radial coordinate $r$ runs from $r=0$, which is the AdS$_5$
horizon, to $r=r_0$, which is a null singularity.  The conformal
factor in these two regions is given by
\bea
r\rightarrow 0:&& e^{2C}\sim \fft{1}{r^{10/3}}\,,\qquad
V\sim \fft{35}{4r^2}\nn\\
r\rightarrow r_0:&& e^{2C}\sim (r-r_0)^{2/9}\,,\qquad
V\sim \fft{1}{12(r-r_0)^2}\,.
\eea
Thus the system has a discrete spectrum, indicating confinement.

\subsection{Lifting to $S^2$-wrapped M5-brane}

   The consistent $S^4$ reduction of eleven-dimensional supergravity
was obtained in \cite{nasvamvan1,lp7,nasvamvan2}.  Using the explicit
reduction ansatz given in \cite{lp7}, we can lift the above solution
to give an $S^2$-wrapped M5-brane in $D=11$, with
\bea
ds_{11}^2\!\!\!&=&\!\!\! \Delta^{1/3}\,\Big(a^2\, dx^\mu\, dx_\mu + b^2\,
(dr^2 + d\Omega_2^2)\Big)+2f^{-1/3}\, 
\Delta^{1/3}\, d\xi^2 + \ft12 \Delta^{-2/3}\, f\,
\cos^2\xi\, (\sigma^2 + d\wtd \Omega_2^2)\,,\nn\\
A_\3\!\!\!&=&\!\!\!\ft{1}{\sqrt2}\sin\xi\, \sigma\, \Big(\wtd\Omega_\2
-\Omega_\2 +\ft12\cos^2\xi\, \Delta^{-1}\, f^4\, \wtd\Omega_\2\Big)\,.
\eea
Here
\be
\Delta=f^4\, \sin^2\xi + f^{-1}\, \cos^2\xi\,,\qquad
d\sigma=\Omega_\2 + \wtd\Omega_\2\,.
\ee
In the asymptotic region at large proper distance, the metric becomes
\be
ds_{11}^2 \sim  \fft{1}{r^2} (dx^\mu\, dx_\mu + d\Omega_2^2 + dr^2)
+ 2 d\xi^2 +  \ft12 \cos^2\xi (\sigma^2 + d\wtd\Omega_2^2)\,,
\ee
where $r\rightarrow 0$.  Note that the 4-form field strength
$F_\4=dA_\3$ has a term
\be
F_\4 = \ft{1}{2\sqrt{2}}\, \Delta^{-1} f^4\, \cos^3(\xi)\, d\xi \wedge
\sigma \wedge \wtd \Omega_2 + \cdots\,,
\ee
implying that this 3-brane configuration has non-vanishing M5-brane
charge.

\subsection{General solutions}

    Although we have not obtained the general solution explicitly, we
can nevertheless show that the first-order equations (\ref{abfeqs})
can be reduced to a single non-linear first-order differential
equation. Defining $X\equiv b/f$, $Y\equiv b\, f^4$, and $dt = 5f d\rho$, we
have
\bea
\fft{a'}{a} = \ft15(-X\, Y - 4 X^2 + 2)\,,\qquad
\fft{X'}{X} =-X\, Y -2\,,\qquad
\fft{Y'}{Y} = 3 X\, Y - 4X^2\,.
\eea
The first equation gives $a$, once $X$ and $Y$ have been found using
the remaining equations. The second
equation may be solved for $Y$, and substituted into the third.  This gives
\be
X\, X'' + {X'}^2 + 4 X^3\, X' + 10 X\, X' + 8X^4 + 12 X^2=0\,.\label{2o}
\ee
Now let $v\equiv X'$, so that $X''= v'= dv/dX\, dX/d\rho = v\,dv/dX$,
and hence (\ref{2o}) becomes
\be
v\, X\, \fft{dv}{dX} + v^2 + 4v\, X^3 + 10 v\, X + 8 X^4 + 12
X^2=0\,.\label{2o2}
\ee
A further change of variable from $v$ to $w$, defined by $w\equiv
\ft12 v\, X$, then gives
\be
X^{-1}\, w\, \fft{dw}{dX} + 2w\,  X^2 + 5 w + 2 X^4 + 3 X^2=0\,.\label{2o3}
\ee
Finally, we let $z\equiv X^2+\ft52$.  This transforms (\ref{2o3}) into
\be
w\, \fft{dw}{dz} + z\, w + \ft12 (z-1)(2z-5) =0\,.\label{2o4}
\ee
This is a particular case of Abel's equation, but
unfortunately it appears to be difficult to obtain the solution in
closed form.

\subsection{Non-abelian solutions in $D=7$ supergravity}

     So far we have made use only of a $U(1)$ subgroup of the $SU(2)$
gauge fiends.  It is possible also to turn on the full
$SU(2)$ gauge fields, with the ansatz
\be
A_\1^1 = v\, \sin\theta\, d\phi\,,\qquad 
A_\1^2 = - v\, d\theta\,,\qquad
A_\1^3 = \cos\theta\, d\phi\,,
\ee
where $v$ is a function of $r$, and $(\theta,\phi)$ are the
coordinates on the 2-spheres foliating the transverse 3-space. 
(This ansatz was used, for example, in
\cite{chamvolk}.)  The Hamiltonian $H=T+V$ for this case is given by
\bea
T&=& -12{A'}^2 - 16 A'\, B' - 2 {B'}^2 + \ft12 {\phi'}^2 +  e^{-2B +
\fft{2}{\sqrt{10}}\,\phi}\, {v'}^2\,,\nn\\
V &=& e^{8A+2B}\, \Big(2 - e^{2B}\, U - \ft12\, e^{-2B + \fft{2}{\sqrt{10}}\,
\phi}\, (v^2-1)^2\Big)\,,
\eea
where $U$ is the scalar potential in $D=7$ gauged supergravity,
as given in (\ref{scalarpot}). 

   We have not found a superpotential for this system.  The earlier
$U(1)$ result corresponds to $v=0$.  There is a singular scaling limit
in which the first two terms in the scalar potential $U$ in
(\ref{scalarpot}) vanish, and then the theory can be viewed as the
$S^3$ reduction of ${\cal N}=1$, $D=10$ supergravity
\cite{chamsab}.  A supersymmetric 3-brane with $SU(2)$ Yang-Mills
fields does then exist, and it is non-singular \cite{maldnunez2}.  The
solution is the lifting to $D=10$ of the $SU(2)$ black hole constructed in
\cite{chamvolk}.  The superpotential for this system was obtained in
\cite{paptsey}.

\section{Conclusions}

    In the context of four-dimensional field theories, it is of
considerable interest to construct 3-brane configurations in M-theory.
One class of such solutions has been obtained by wrapping M5-branes on
a certain supersymmetric two cycles, such as Riemannian surfaces
\cite{maldnunez1}.  These solutions are deformations of an
AdS$_5\times H^2\times S^4$ vacuum, where $H^2$ is the hyperbolic
plane.

          In this paper, we have constructed two new types of 3-brane
configuration.  In the first type, we exploit the fact that the
transverse space of the 3-brane in $D=11$ is seven-dimensional, and
that there exist non-trivial seven-dimensional Ricci-flat manifolds
with $G_2$ holonomy.  It has been proposed that compactifications of 
M-theory on $G_2$ are related to ${\cal N}=1$, $D=4$ Yang-Mills theory
\cite{acharya,amv,wit-talk}.  Three explicit complete 
non-compact manifolds of $G_2$ holonomy are currently known 
\cite{brysal,gibpagpop}, and they can be used to smooth out the
singularities of compact $G_2$ orbifolds.  Each of them admits a 
harmonic 4-form \cite{clptrans,d2ns2}, which suggests the possibility
of turning on the 4-form field strength of $D=11$ supergravity in
solutions that correspond to deformations of four-dimensional
Minkowski space times a Ricci-flat $G_2$ manifold.  We have indeed
managed to obtain exact solutions of this type, which can be viewed as
3-branes in M-theory.  

    The general solutions contain two continuous parameters; $\ell$,
which measures the size of the gravitional instanton, and $c$, which
measures the strength of the 4-form.  The solutions are massless, and
carry no 4-form charge.  In common with all other known massless brane
solutions, there are naked singularities at short distance.  At large
distance the solution approaches the product of four-dimensional
Minkowski spacetime and the original Ricci-flat $G_2$ manifold.
The solution can be viewed as a supergravitional monopole, involving both
the metric and the 4-form in the supergravity multiplet.  

    We obtained the solutions by deriving first-order equations from a
superpotential, and we showed that these are precisely the
integrability conditions for the existence of a Killing spinor.  Thus
the two M3-brane solutions that we have constructed in this paper are
supersymmetric.

    It is interesting to observe that although the harmonic 4-forms in
the undeformed manifolds of $G_2$ holonomy, on the $\R^3$ bundle over
$S^4$ and the $\R^4$ bundle over $S^3$, have quite different
properties (the former being $L^2$ normalisable whilst the latter is
not), the corresponding deformed 3-brane solutions in sections 2 and 3
have very similar qualitative behaviour.  In contrast, the properties
of the metrics for the fractional D2-brane \cite{d2ns2} and NS-NS
2-brane \cite{clptrans}, which make use of these same two Ricci-flat
metrics, are significantly different. 

    We also obtained M3-brane solutions of a different kind, by
lifting 3-brane solutions in $D=7$ gauged supergravity back to
$D=11$.  They carry magnetic 4-form charge, and 
can be viewed as M5-branes wrapped on $S^2$.  

\section*{Acknowledgement}

       We are grateful to Mike Duff, Gary Gibbons, James Liu and
Jianxin Lu for useful discussions.  C.N.P. is grateful to the Michigan
Center for Theoretical Physics for hospitality during the completion
of this work.

\section*{Note added}

    In an earlier version of this paper it was claimed that the
M3-brane solutions were not supersymmetric, but instead were
``pseudo-supersymmetric'' with respect to a modified $D=11$
supersymmetry transformation rule.  This incorrect conclusion resulted
from a systematic error in a computer program that we used for
calculating the Killing spinors.  We are grateful to Jim Liu for
calculations that encouraged us to recheck the computer programs and
discover the error.

\end{document}